\documentclass[12pt,a4paper]{article}
\usepackage{amssymb,amsfonts,amsmath,amsthm}
\usepackage[cp1251]{inputenc}
\usepackage[russian]{babel}
\usepackage{graphicx}
\usepackage{enumerate}
\allowdisplaybreaks   

\begin{document}
\pagestyle{empty}

ЛЕОНЕНКО Василий Николаевич

РОССИЯ, 644099, г. Омск

ул.Певцова, 13

Омский филиал Института математики им. С. Л. Соболева СО РАН

e-mail: VNLeonenko@yandex.ru

УДК 519.711
\begin{center} В. Н. Леоненко
\\
\vspace{\baselineskip} \textbf{ИНДИВИДУУМ--ОРИЕНТИРОВАННАЯ МОДЕЛЬ РАСПРОСТРАНЕНИЯ ИНФЕКЦИОННОГО
ЗАБОЛЕВАНИЯ \\ В ГОРОДСКОЙ СРЕДЕ}
\end{center}
\vspace{\baselineskip}

{\small Рассматривается индивидуум--ориентированная модель распространения инфекционного заболевания среди
городского населения. Построена система стохастических уравнений, описывающих изменение численностей четырёх
групп населения: восприимчивые, инфицированные, больные индивидуумы и индивидуумы, находящиеся в состоянии
ремиссии. Система уравнений модели дополнена соотношениями, учитывающими тяжесть и продолжительность болезни для
каждого заболевшего индивидуума. На основе методов Монте--Карло разработаны алгоритм и моделирующая программа,
позволяющие исследовать динамику численностей указанных групп населения.}

{\small An individual-based model of the infectious disease spread among the urban population is considered. A
system of stochastic equations, which describes changes in quantities of four population groups, susceptible,
exposed, infected individuals and individuals in the state of remission, is built. The system of equations of
the model is supplemented with correlations, which consider disease heaviness and duration for every infected
individual. An algorithm and a modelling program based on Monte--Carlo methods which allows to investigate the
group number dynamics is developed.}

\begin{center}{\bf Введение}\end{center}
Одним из современных направлений математического моделирования является изучение закономерностей
социально--демографических и эпидемических процессов, анализ и контроль распространения инфекционных
заболеваний, а также выработка различных мер по их сдерживанию \cite{survey}, \cite{may}. Для решения
поставленных задач применяется достаточно широкий класс математических моделей, среди которых особое место
занимают индивидуум--ориентированные модели \cite{survey}, \cite{prk}, \cite{levdur}. В настоящей работе
рассматривается имитационная модель распространения инфекционного заболевания в городской среде, построенная в
виде системы стохастических рекуррентных уравнений. Описана модификация модели, учитывающая особенности
протекания болезни у заражённых индивидуумов. Приводятся результаты вычислительного эксперимента с
использованием численных методов Монте--Карло.

\begin{center}{\bf Описание модели}\end{center}

Считаем, что население некоторого города состоит из четырёх групп: восприимчивые к инфекции индивидуумы (S),
инфицированные индивидуумы (L), больные индивидуумы (T), индивидуумы в состоянии ремиссии (R). Время модели
является дискретным, с единичным шагом. Пусть $x_{t}$ -- численность индивидуумов из S, $y_{t}$ -- численность
индивидуумов из L, $z_{t}$ -- численность индивидуумов из T, $r_{t}$ -- численность индивидуумов из R в моменты
времени $t=0,1,2,\ldots.$ В течение промежутка $(t,t+1]$ для индивидуумов указанных групп населения возможны
следующие события:
\renewcommand{\labelitemi}{}
\begin{itemize}
\item А1) гибель индивидуума из S с вероятностью $1-\gamma$, \item А2) гибель индивидуума из L с вероятностью
$1-\gamma$, \item А3) гибель индивидуума из T с вероятностью $1-\rho$, \item А4) гибель индивидуума
из R с вероятностью $1-\sigma$, \item А5) инфицирование индивидуума из S в результате контакта с
больным индивидуумом (переход из S в L),\item А6) активация болезни у индивидуума из L в результате
повторного инфицирования -- контакта с больным индивидуумом (переход из L в T),\item А7) спонтанная
активация болезни у индивидуума из L с вероятностью $\beta$ (переход из L в T),\item А8) спонтанное
выздоровление индивидуума из T с вероятностью $\alpha$ (переход из T в R), \item А9) активация
болезни у индивидуума из R в результате повторного инфицирования -- контакта с больным индивидуумом
(переход из R в T), \item А10) спонтанная активация болезни у индивидуума из R с вероятностью
$\theta$ (переход из R в T). \end{itemize} Полагаем, что вероятности $\gamma, \rho, \sigma, \beta,
\alpha, \theta$ удовлетворяют следующим ограничениям: $\gamma \in(0;1)$, $\rho \in(0;\gamma]$,
$\sigma \in [\rho, \gamma]$, $\beta \in(0;1)$, $\alpha\in(0;1)$, $\theta\in(0;1)$. Для описания
процессов инфицирования индивидуумов и активации болезни в результате повторного инфицирования
используется схема, предложенная в работе \cite{prk}. Считается, что за единицу времени больные
индивидуумы посещают в совокупности $\xi_t$ мест, где $\xi_t = \sum_{i=1}^{z_t} \psi_{it}$,
$\psi_{it} \geq 0$ -- взаимно независимые, одинаково распределённые случайные величины, не
зависящие от $z_t$, с математическим ожиданием $E(\psi_{it}) = \pi >0$. Вероятности того, что
индивидуумы из групп S, L или R посетят одно из этих мест, обозначим через $\lambda_S\in(0;1)$,
$\lambda_L\in(0;1)$ и $\lambda_R\in(0;1)$ соответственно. Вероятности заражения индивидуумов после
контакта с больным индивидуумом обозначим через $\delta_S\in(0;1)$, $\delta_L\in(0;1)$ и
$\delta_R\in(0;1)$ соответственно. Тогда вероятности того, что индивидуум из конкретной группы S,
L, R будет инфицирован в течение суток (с учётом возможности спонтанной активации болезни для
индивидуумов групп L и R), при фиксированном значении $\xi_t$ задаются следующими формулами:
$\mu_t^{(S)} = 1 - (1-\lambda_S \, \delta_S)^{\xi_t}$, $\, \mu_t^{(L)} = 1 - (1-\beta)\,
(1-\lambda_L \, \delta_L)^{\xi_t}$, $\, \mu_t^{(R)} = 1 - (1-\theta)\, (1-\lambda_R \,
\delta_R)^{\xi_t}$. Наряду с событиями А1) -- А10) будем учитывать процессы миграции, которые могут
оказывать значительное влияние на изменение численностей изучаемых групп населения. Считаем, что
каждая группа S, L, T, R пополняется за промежуток времени $(t,t+1]$ в количестве $f_{t+1}\geq 0$,
$\, g_{t+1}\geq 0$, $\, h_{t+1} \geq 0$, $\, d_{t+1}\geq 0$ индивидуумов. Cистема уравнений на
численности индивидумов имеет вид:
\begin{gather}
x_{t+1} = \widehat{x}_{t+1} - u_{t+1} + f_{t+1}, \nonumber \\
\label{system} y_{t+1} = \widehat{y}_{t+1} + u_{t+1} - w_{t+1} + g_{t+1}, \\
z_{t+1} = \widehat{z}_{t+1} - v_{t+1} + w_{t+1} + l_{t+1} + h_{t+1},  \nonumber \\
r_{t+1} = \widehat{r}_{t+1} + v_{t+1} - l_{t+1} + d_{t+1}, \quad t = 0,1,2,\ldots,  \nonumber \\
x_0 = x^{(0)}\geq 0, \, y_0 = y^{(0)}\geq 0, \, z_0 = z^{(0)}\geq 0, \, r_0 = r^{(0)}\geq 0.
\nonumber
\end{gather}
Здесь $\widehat{x}_{t+1}$, $\widehat{y}_{t+1}$, $\widehat{z}_{t+1}$, $\widehat{r}_{t+1}$ -- количество
индивидуумов $x\in S$, $y \in L$, $z \in T$, $r \in R$, доживших от момента времени t до момента t+1; $u_{t+1}$
-- количество индивидуумов из S, подвергшихся инфицированию за время $(t,t+1]$, $w_{t+1}$ -- количество
индивидуумов из L, заболевших за время $(t,t+1]$, $l_{t+1}$ -- количество индивидуумов из R, заболевших за время
$(t,t+1]$, $v_{t+1}$ -- количество индивидуумов из T, выздоровевших за время $(t,t+1]$. Считаем, что все
индивидуумы ведут себя независимо друг от друга. Тогда для заданного $t$ и при фиксированных значениях
соответствующих случайных величин указанные слагаемые имеют условные биномиальные распределения:
$$
\widehat{x}_{t+1} \sim Bin(x_t, \gamma), \, \widehat{y}_{t+1} \sim Bin(y_t, \gamma),
\,\widehat{z}_{t+1} \sim Bin(z_t, \rho),
$$
$$
\widehat{r}_{t+1} \sim Bin(r_t, \sigma), \, u_{t+1} \sim Bin(\widehat{x}_{t+1}, \mu_t^{(S)}), \,
w_{t+1} \sim Bin(\widehat{y}_{t+1}, \mu_t^{(L)}),
$$
$$
\, v_{t+1} \sim Bin(\widehat{z}_{t+1}, \alpha), l_{t+1} \sim Bin(\widehat{r}_{t+1}, \mu_t^{(R)}).
$$

Исследование свойств решений модели (\ref{system}) представляет собой достаточно трудную задачу.
Рассмотрим здесь один из подходов к анализу математических ожиданий переменных модели: $m_t =
E(x_t)$, $n_t = E(y_t)$, $k_t = E(z_t)$, $s_t = E(r_t)$. Без ограничения общности примем, что
начальные значения в (\ref{system}) являются заданными константами. Предположим, что математические
ожидания миграционных потоков являются ограниченными: $\overline{f}_t = E(f_t)\leq \overline{f}$,
$\overline{g}_t = E(g_t) \leq \overline{g}$, $\overline{h}_t = E(h_t) \leq \overline{h}$,
$\overline{d}_t = E(d_t) \leq \overline{d}$, где $\overline{f}$, $\overline{g}$, $\overline{h}$,
$\overline{d}$ -- некоторые константы. Положим $\overline{c} = \overline{f} + \overline{g} +
\overline{h} + \overline{d}$. Используя систему (\ref{system}), получаем
$$
\begin{array}{l}
m_{t+1} + n_{t+1} + k_{t+1} + s_{t+1} = E(\widehat{x}_{t+1} + \widehat{y}_{t+1}+
\widehat{z}_{t+1}+\widehat{r}_{t+1} + f_{t+1} + g_{t+1} + h_{t+1} + d_{t+1}) \\
= \gamma m_t+\gamma n_t+\rho k_t + \sigma s_t + (\overline{f}_{t+1} + \overline{g}_{t+1} +
\overline{h}_{t+1} + \overline{d}_{t+1}) \leq \gamma (m_t + n_t + k_t + s_t) + \overline{c}.
\end{array}
$$
Отсюда следует, что математическое ожидание общей численности населения ограничено. Для оценки
численностей отдельных групп населения можно применить подход, описанный в \cite{prk}, \cite{kas}.
Нетрудно заметить, что при ${m_0 = x^{(0)} \leq \overline{f}/(1-\gamma) = \overline{m}}$ для любого
$t \geq 0$ выполняется неравенство $m_t \leq \overline{m}$. Для переменных $n_t, k_t, s_t$
выписывается система сравнения:
\begin{gather*} x^{(1)}_{t+1} = \gamma (1-\beta) x^{(1)}_t + \gamma \overline{m}(1-(1-\lambda_S\delta_S)^{\pi
x^{(2)}_t}) +
\overline{g}_{t+1}, \\
\begin{array}{l} x^{(2)}_{t+1} = \rho (1-\alpha)x^{(2)}_t + \gamma x^{(1)}_t (1-(1-\beta)(1-\lambda_L\delta_L)^{\pi x^{(2)}_t}) + \\
\qquad {} + \sigma x^{(3)}_t (1-(1-\theta)(1-\lambda_R\delta_R)^{\pi x^{(2)}_t}) + \overline{h}_{t+1}, \end{array}\\
x^{(3)}_{t+1} = \sigma (1-\theta) x^{(3)}_t + \alpha \rho x^{(2)}_t + \overline{d}_{t+1}, \quad t=0,1,2, \ldots,  \\
x^{(1)}_0 = y^{(0)}, \, x^{(2)}_0 = z^{(0)}, \, x^{(3)}_0 = r^{(0)}.
\end{gather*}
Решения этой системы дают оценки сверху на математические ожидания решений системы (\ref{system}).
Эти оценки позволяют получать предварительные выводы о динамике численностей групп, не прибегая к
моделированию методом Монте--Карло.

Рассмотрим модификацию модели, в которой для каждого больного учитывается текущая тяжесть болезни $h = h(t) \geq
0$. Используя \cite{march1}, полагаем, что $ h(t) = h(0)\, \exp(-a\, t + b\, \omega(t)), \ \ t \geq 0$. Здесь:
$h(0) \geq 0$ -- начальное значение тяжести, $a
> 0$, $b > 0$ -- параметры, $\omega(t)$ -- стандартный винеровский случайный процесс. Принимаем,
что $h(0) \sim Unif[0;1]$, то есть начальная тяжесть равномерно распределена на отрезке $[0;1]$, $a$ --
некоторая константа, одинаковая для всех больных, $b\sim Unif[0; \sqrt{2a}]$ -- индивидуальный параметр,
отражающий зависимость тяжести от различных факторов. Границы изменения $b$ выбирались исходя из необходимости
выполнения неравенства $a > b^2/2$, которое обеспечивает уменьшение математического ожидания тяжести болезни,
$E(h(t))$, с течением времени. Примем далее, что параметры $\rho$, $\alpha$ и $r$ зависят от тяжести болезни и
тем самым имеют различные значения для разных больных. Полагая, что больной с нулевой тяжестью идентичен
здоровому индивидууму, законы изменения указанных величин относительно $h$ задаём следующим образом: ${\rho(h) =
\gamma\, \exp(-\varepsilon_1 \, h)}$, ${\alpha(h) = \exp(-\varepsilon_2 \, h)}$, ${r(h) = r_0 \,
\exp(-\varepsilon_3 \, h)}$, где $\varepsilon_1>0$, $\varepsilon_2>0$, $\varepsilon_3>0$ -- параметры модели.
Опираясь на это предположение, приходим к тому, что вид слагаемых $\widehat{y}_{t+1}$ и $v_{t+1}$ в системе
(\ref{system}) изменяется. Пусть $h_{it}$ -- тяжесть заболевания i--го индивидуума из T в момент времени t.
Тогда $\widehat{y}_{t+1} = \sum_{i=1}^{y_t}\xi_{it}$, $\, v_{t+1} = \sum_{i=1}^{\widehat{z}_{t+1}}\eta_{it}$,
где
$$
 \mbox{ } \xi_{it} = \left\{\begin{array}{l} 1, \mbox{с вер. }\rho(h_{it}), \\ 0, \mbox{с вер. }
1-\rho(h_{it}); \end{array}\right. \mbox{ } \eta_{it} = \left\{\begin{array}{l} 1, \mbox{с вер. }\alpha(h_{it}),
\\ 0, \mbox{с вер. } 1-\alpha(h_{it}). \end{array}\right.$$

Отметим, что величины $\psi_{it}$, задающие $\xi_t$, будут иметь различные распределения, поскольку
$\pi\neq const$.

\begin{center}{\bf Результаты вычислительных экспериментов}\end{center}

В таблице приводятся доверительные интервалы для математических ожиданий $m_t$, $n_t$, $k_t$, $s_t$ численностей
групп базовой модели (модель 1) и модифицированной модели (модель 2) в момент времени $t = 200$ на уровне
доверия 0.95 (объём выборки n=100). Начальные численности: $x^{(0)} = 100000$, $y^{(0)} = 50000$, $z^{(0)} =
500$, $r^{(0)} = 40$. Значения входных потоков: $f_t \equiv 1000.0$, $g_t \equiv 0.0$, $h_t \equiv 10.0$, $d_t
\equiv 0.0$.

\vspace{0.5cm}
\small
\begin{tabular}{|c|c|c|c|c|c|}
\hline
Эксп. & Мод. & $m_t$ & $n_t$ & $k_t$ & $s_t$\\
\hline
$1$ & $1$ & $99886.6\pm 63.43$ & $6707.79 \pm 15.11 $& $59.47 \pm 1.55 $& $7.94 \pm 0.51$ \\
$1$ & $2$ & $99901.6\pm 55.08$ & $6707.7 \pm 16.29 $& $54.57 \pm 1.3 $& $17.14 \pm 0.86$ \\\hline
$2$ & $1$ & $65483.4 \pm 52.01$ & $1513.79 \pm 8.04$ & $30267.3 \pm 51.95$ & $726.12 \pm 4.95$ \\
$2$ & $2$ & $98715.1\pm 53.82$ & $30.94 \pm 0.97 $& $434.96 \pm 4.2 $& $29.23 \pm 1.05$ \\\hline
\end{tabular}
\normalsize

\vspace{0.5cm}

В случае эксперимента 1 инфекция действует на группу S незначительно. Численность больных индивидуумов сразу
начинает падать, а затем стабилизируется на низком уровне. Эксперимент 2 иллюстрирует следующую ситуацию:
численность больных индивидуумов на начальном этапе быстро возрастает до больших значений (максимальное значение
$k_t = 45306.5 \pm 29.4425$ при $t=35$), а затем начинает медленно падать. При этом инфицированию подвергается
значительная часть индивидуумов группы S.

Для сравнения приводятся данные по экспериментам 1,2 с использованием модели 2. Значения
параметров, общих для обеих моделей, брались из соответствующих наборов для экспериментов модели 1.
Остальные параметры принимали следующие значения: $a = 0.12$, $\varepsilon_1 = \varepsilon_2 =
\varepsilon_3 = 2.0$. Динамика численностей групп заметно отличается в реализациях на базе разных
моделей. Особенно это заметно в случае эксперимента 2: возрастание количества больных в
модифицированной модели стало гораздо более кратковременным, а максимальная численность больных
существенно уменьшилась. В результате группа S оказалась затронута инфекцией менее значительно.

В заключение отметим, что описанный подход может быть использован для построения стохастической
индивидуум--ориентированной модели распространения туберкулеза \cite{march2}.

Автор благодарит своего научного руководителя профессора Н. В. Перцева (ОФ ИМ им. С. Л. Соболева СО РАН) за
постановку задачи и обсуждение результатов работы.

 \renewcommand{\refname}{\begin{center} \normalsize Список литературы \end{center}}

\end{document}